\documentclass[twocolumn]{webofc}

\usepackage[varg]{txfonts}   
\begin{document}
\title{Role of charge equilibration in multinucleon transfer in damped collisions of heavy ions}
%
%

\author{\firstname{Vyacheslav} \lastname{Saiko}\inst{1,2}\fnsep\thanks{\email{saiko@jinr.ru}}
        \and
        \firstname{Alexander} \lastname{Karpov}\inst{1,2}\fnsep\thanks{\email{karpov@jinr.ru}}
}

\institute{Flerov Laboratory of Nuclear Reactions, JINR, Dubna, Russia
\and
           Dubna State University, Dubna, Russia
}

\abstract{%
In this work the charge equilibration process has been analyzed within the Langevin-type dynamical approach. Its duration and energy dependence is discussed.
We have analyzed the isotopic distributions of final products obtained in the isospin-asymmetric $^{58}$Ni,$^{40}$Ca + $^{208}$Pb reactions. Comparison of  $^{58}$Ni,$^{64}$Ni + $^{208}$Pb systems have been done in order to analyze the final yields of neutron-rich heavy nuclides.
}
\maketitle
\section{Introduction}
\label{intro}

Multinucleon transfer (MNT) is one of the main processes observed in collisions of heavy ions at energies near the Coulomb barrier. It leads to the production of exotic nuclei formed in channels with a transfer of dozens of nucleons between projectile and target. Nowadays this type of reaction is considered as a promising method of production of neutron-rich isotopes of heavy elements hardly achievable by other reactions, e.g. fusion or fragmentation. These nuclides and their properties are crucial for detailed understanding the pathway of the astrophysical r-process.

Langevin-type dynamical models are powerful tools for analyzing the processes of low-energy nuclear collisions (e.g. fusion, fission, deep-inelastic scattering)~\cite{Frobrich,Karpov,Zagrebaev}. They provide an appropriate macroscopic description of various measurable characteristics such as angular, mass, and energy distributions of reaction products by taking into consideration the concept of friction and the statistical fluctuations in collisions of heavy nuclei.

One of such the models has been recently developed and has successfully described a bulk of experimental data obtained in the following MNT reactions: $^{136}$Xe + $^{198}$Pt,$^{208}$Pb,$^{209}$Bi; $^{160}$Gd + $^{186}$W; $^{208}$Pb + $^{208}$Pb; $^{238}$U + $^{238}$U,$^{248}$Cm using the same set of the model parameters~\cite{KarpovSaiko,SaikoKarpov}. All these combinations are characterized by close values of neutron-to-proton ratio $N/Z$ of projectiles and targets and can be called as isospin-symmetric ones. In the present work we aimed at analysis of isospin-asymmetric systems consisted of nuclei with quite a different $N/Z$: $^{40}$Ca,$^{58}$Ni + $^{208}$Pb (see Table~\ref{tabNZ}). In this case, the process of reaching an equilibrium of the $N/Z$ value during collision usually called as the charge equilibration (or isospin transport) plays a significant role in reaction dynamics and formation of the fragments.

\begin{table}[b]
\centering
\caption{$N/Z$ values of projectile and target for the studied reactions}
\label{tabNZ}
\begin{tabular}{|l|l|l|}
\hline
Combination & $(N/Z)_{projectile}$ & $(N/Z)_{target}$ \\\hline \hline
$^{136}$Xe + $^{198}$Pt & 1.519 & 1.538 \\\hline
$^{136}$Xe + $^{208}$Pb & 1.519 & 1.537 \\\hline
$^{136}$Xe + $^{209}$Bi & 1.519 & 1.518 \\\hline
$^{160}$Gd + $^{186}$W & 1.5 & 1.514 \\\hline
$^{238}$U + $^{238}$Cm & 1.587 & 1.583 \\\hline
$^{64}$Ni + $^{208}$Pb & 1.286 & 1.537 \\\hline
$^{58}$Ni + $^{208}$Pb & 1.071 & 1.537 \\\hline
$^{40}$Ca + $^{208}$Pb & 1 & 1.537 \\\hline
\end{tabular}
\end{table}

\section{Model}
\label{Model}

A collision of two heavy nuclei is considered in the model in a continuous way as a three-step process. At the first stage, nuclei approach each other, then they form a mononuclear system, which evolves and finally decays into two fragments. The evolution of the system of colliding nuclei is governed by the multidimensional potential energy calculated within the macro-microscopic approach~\cite{ZagrebaevKarpov}, based on the two-center shell model (TCSM)~\cite{MaruhnGreiner}. We use eight degrees of freedom in the model, four of them originate from the TCSM parametrization and define nuclear shapes: elongation, two ellipsoidal deformations, and mass asymmetry. The charge asymmetry $\eta_Z=(Z_2-Z_1)/(Z_1+Z_2)$ is the variable defining the $N/Z$ ratio of fragments (at fixed masses) and, therefore, plays the most important role in describing the charge equilibration process, which is the topic of the article.

Evolution of all the degrees of freedom is treated by the set of the Langevin equations:
\begin{eqnarray}\label{LangevinEqs}
&&\dot q_i = \sum_j \mu_{ij} p_j, \\
&&\dot p_i = F_i^{\rm driving} + F_i^{\rm friction} + F_i^{\rm random}. \nonumber
\end{eqnarray}
Here $\mu_{ij}=\|m_{ij}\|^{-1}$ are the inverse mass coefficients calculated within the Werner-Wheeler hydrodynamical approach~\cite{WW}, $q_i$ and $p_i$ are degrees of freedom and their momenta, respectively. The driving force is given mainly by a derivative of the potential energy. The friction coefficients $\gamma_{ij}$ determining the friction force are calculated using the wall-and-window model of one-body dissipation~\cite{one-body}. According to the Einstein's relation, the amplitude of the random force is proportional to $\sqrt{\gamma_{ij}T}$, where $T$ is the nuclear temperature.

The system of Langevin equations~(\ref{LangevinEqs}) is solved numerically with the following initial conditions: two nuclei start to approach each other from a distance $\approx50$ fm with a given impact parameter and collision energy in the center-of-mass frame. Calculation of a trajectory is terminated when the distance between formed fragments reaches initial value. Thus, the obtained trajectory keeps all characteristics of the collision and primary products, such as nucleon numbers, scattering angles, kinetic and excitation energies, etc.

All these characteristics are the input data for the statistical model, which is used to simulate a decay cascade of excited primary products in order to obtain the final ones~\cite{stat}. The emission of neutrons, protons, $\alpha$-particles, and $\gamma$-quanta as well as sequential fission process are considered in the calculations. In the latter case, masses and atomic numbers of the sequential fission fragments are simulated using the GEF code~\cite{GEF}. Thus, the calculation results can be directly compared with the experimentally measured data.

Employing the described approach, we simulate a large number of trajectories for each relevant impact parameter and get the differential cross section of various processes in the standard way:
\begin{equation}\label{CS}
\frac{d^4\sigma}{dZdAdEd\Omega}(Z,A,E,\theta)= \int\limits_0^{b_{\rm max}} \frac{\Delta N(b,Z,A,E,\theta){b\,db}}{N_{\rm tot}(b){\Delta Z \Delta A \Delta E \sin{\theta} \Delta \theta}},
\end{equation}
where $\Delta N(b,Z,A,E,\theta)$ is the number of events in a given bin and $N_{\rm tot}(b)$ is the total number of events simulated for each impact parameter. Integration of~(\ref{CS}) allows one to obtain various distributions of reaction products. The detailed description of the dynamical and statistical approaches mentioned above can be found in~\cite{KarpovSaiko}.

\section{Results and discussion}
\label{Results}

\begin{figure*}[ht]
\centering
\includegraphics[width=1\linewidth]{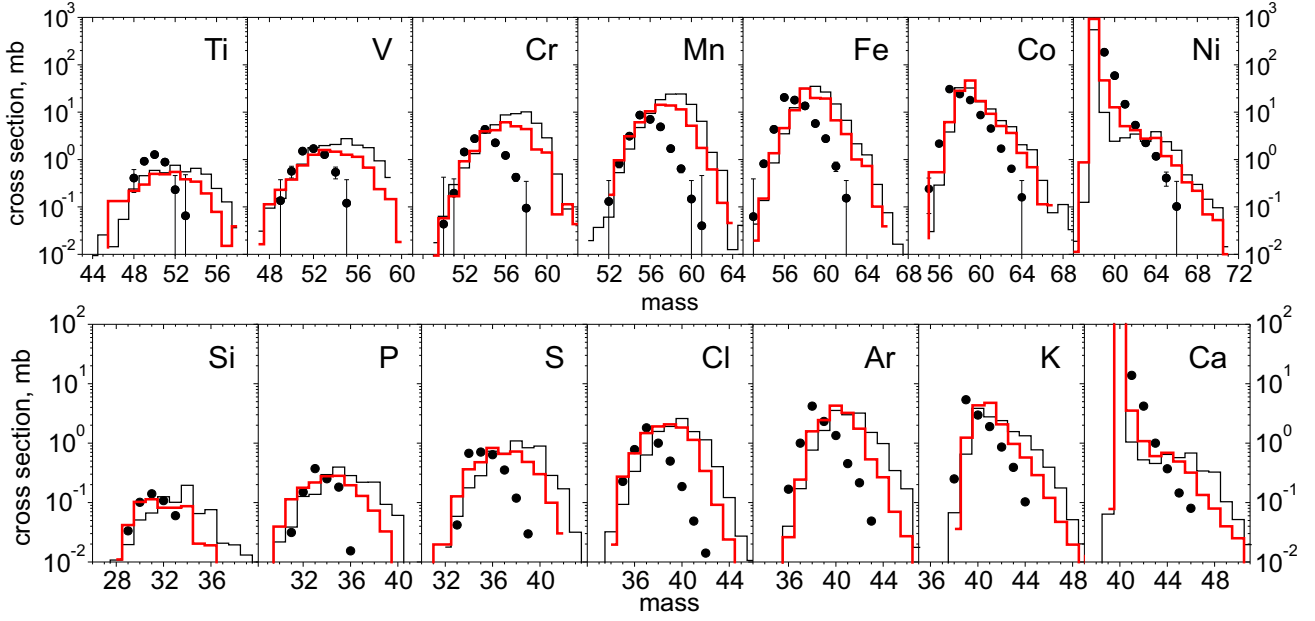}
\caption{Final isotopic distributions of PLFs obtained in the $^{58}$Ni + $^{208}$Pb reaction at $E_{\rm lab.}=328.4$ MeV (upper panel) and in the $^{40}$Ca + $^{208}$Pb reaction at $E_{\rm lab.}=249$ MeV (bottom panel). Theoretical results obtained before and after the model improvement are shown by the thin and thick histograms, respectively. Symbols indicate the experimental data taken from~\cite{CorradiNi,SzilnerCa}.}
\label{img:Ni58Ca40}
\end{figure*}

In spite of the fact that the model describes rather well the MNT for many systems (see~\cite{KarpovSaiko,SaikoKarpov}), it fails to reproduce the experimental data for isospin-asymmetric combinations accurately enough. Particular example is the isotopic distributions of final projectile-like fragments (PLFs) obtained in the $^{40}$Ca,$^{58}$Ni + $^{208}$Pb reactions ($E_{\rm lab.}($Ca$)=249$ MeV and $E_{\rm lab.}($Ni$)=328.4$ MeV, respectively) shown in Fig.~\ref{img:Ni58Ca40} by thin histograms. The angular ranges $42^{\rm o}<\theta_{\rm lab.}<100^{\rm o}$ and $70^{\rm o}<\theta_{\rm lab.}<110^{\rm o}$ covered in the corresponding experiments~\cite{SzilnerCa,CorradiNi} had been taken into account in the calculations. First, the heavy (neutron-rich) part of the isotopic distributions is overestimated for all values of $Z$ in Fig.~\ref{img:Ni58Ca40}. It is essential to rectify this failure in order to provide reasonable predictions of yields of neutron-rich nuclides, which are a matter of priority. Second, the calculated shapes of the Ca and Ni distributions contain a plateau, while the experimental data indicate almost an exponential trend. Mainly it is due to the pickup channels of 1-3 neutrons are strongly underestimated by the model.

The nucleon transfer is described as a collective process in the model, while the mechanism of few nucleon transfer might be a single-particle one. Therefore, calculated cross sections can significantly differ from the experimental data.


We have found that calculated isotopic distributions of target-like fragments (TLFs) obtained in mass-asymmetric reactions (not shown) have been shifted relative to the experimental data, but in the opposite direction than for PLFs. It led us to reconsideration of the division of the excitation energy between two fragments. Thus, instead of fully mass-proportional sharing of the energy, we assume the equal partition at the early stage, since nearly the same number of nucleons in a projectile and a target are initially involved in the dissipation process. Experimental investigation of mass-asymmetric collisions~\cite{Awes,Vandenbosch}, as well as microscopic calculations~\cite{Umar}, have confirmed this assumption. We treat the transition from the equal division of the excitation energy to the mass-proportional one as
\begin{equation}\label{Exc}
E^*_i= \frac{E^*}{2}\exp\left(-t/\tau\right) + \frac{A_i}{A_{tot}}E^*\left[1-\exp\left(-t/\tau\right)\right],
\end{equation}
with the characteristic transition time $\tau$=2 zs. Here $E^*$ is the total excitation energy of a nuclear system, $t$ is the interaction time, $A_{tot}$ and $A_i$ are the total and the fragment masses, respectively.

This model modification leads to increase of the excitation energy of the lighter fragments and, thus, to increase of the number of evaporated neutrons. The corresponding isotopic distributions of final PLFs shift to lighter masses. The trend is obviously opposite for the TLFs: their lower excitation energy leads to isotopic distributions shifted toward heavier masses. Thus, this correction favors the survival of neutron-rich isotopes of heavier elements in the mass-asymmetric reactions.

The problem with describing the few-neutron pickup channels has been solved efficiently by the twice increased random force for the charge-asymmetry degree of freedom at the initial stage of a collision. Calculations showed that this regime of doubled fluctuations of the charge asymmetry is appropriate until the excitation energy reaches the value of 10 MeV.

The results obtained within the modified model are shown in Fig.~\ref{img:Ni58Ca40} by thick histograms. Overall agreement with the experimental data has been improved. The adopted corrections have improved the shape of the Ni and Ca distributions. It has become smoother and consistent with the experimental data, however, a small deviation from the data is still presented in the calculations. The shift of the centroids of the calculated isotopic distributions toward the corresponding experimental values is negligible for the few-proton-transfer channels. However, it increases with the number of transferred protons reaching the experimental data.

\begin{figure}[t]
\centering
\includegraphics[width=8 cm,clip]{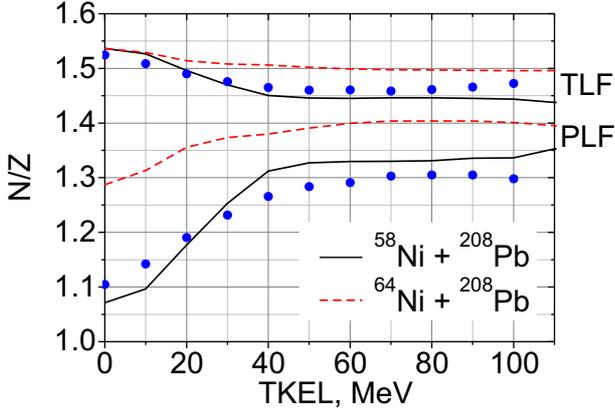}
\caption{Charge equilibration process in the $^{58}$Ni,$^{64}$Ni + $^{208}$Pb reactions as a function of total kinetic energy losses (TKEL). Symbols are the experimental data obtained in~\cite{SapottaNi} for the $^{58}$Ni-induced reaction.}
\label{img:N/Z_TKEL}
\end{figure}

Further, we have considered the charge equilibration process in the studied systems. First, it occurs via the proton and neutron transfer in opposite directions as following the derivative of the potential energy with respect to the mass and charge asymmetries. In particular, the dependence of the $N/Z$ equilibration on the TKEL is shown in Fig.~\ref{img:N/Z_TKEL}. The theoretical curves for the $^{58}$Ni + $^{208}$Pb system are consistent with the experimental data taken from~\cite{SapottaNi}. According to the calculations, the charge equilibration process in the considered systems takes 0.5 zs, which coincides with the corresponding relaxation time obtained within the microscopic TDHF model~\cite{Umar}.

\begin{figure*}[ht]
\centering
\includegraphics[width=1\linewidth]{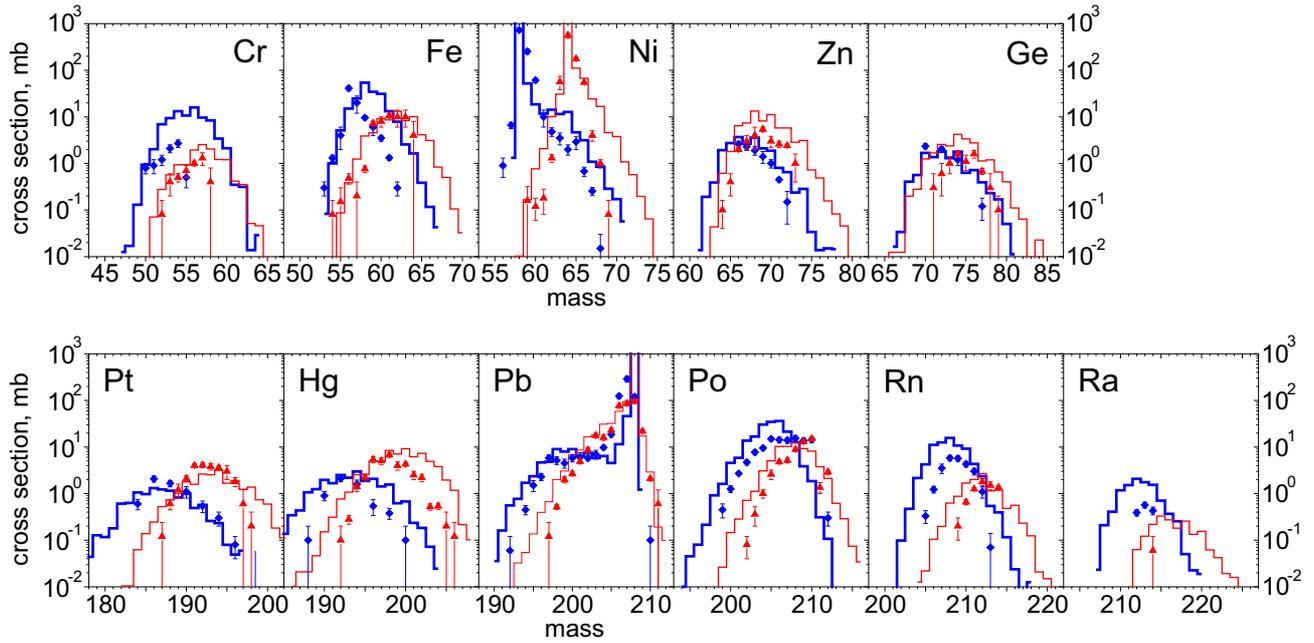}
\caption{Final isotopic distributions of PLFs and TLFs obtained in two reactions: $^{58}$Ni + $^{208}$Pb, $E_{\rm lab.}=345$ MeV (diamonds and thick histograms) and $^{64}$Ni + $^{208}$Pb, $E_{\rm lab.}=350$ MeV (triangles and thin histograms). The histograms are the results of improved calculations and the symbols are the experimental data from~\cite{KrolasNi58,KrolasNi64}.}
\label{img:Ni5864}
\end{figure*}

Finally, the $^{58}$Ni,$^{64}$Ni + $^{208}$Pb reactions ($E_{\rm lab.}=345, 350$ MeV) have been investigated. The final isotopic distributions for Z-even elements are shown in Fig.~\ref{img:Ni5864} in  comparison with the data from Ref.~\cite{KrolasNi58,KrolasNi64}.

The charge equilibration process taking place in the isospin-asymmetric $^{58}$Ni + $^{208}$Pb reaction favours larger yields of lightest PLFs compared to the isospin-symmetric $^{64}$Ni + $^{208}$Pb reaction. Comparing the two reactions, one can notice that the final distributions of PLFs become closer to each other with larger nucleon transfer, while for TLFs the situation is vice versa. It makes the production of neutron-rich isotopes of heavy elements more preferable in the isospin-symmetric $^{64}$Ni + $^{208}$Pb system, which has a larger neutron excess.

In conclusion, the charge equilibration process has been analyzed within the Langevin-type dynamical approach. This process is governed by the potential energy landscape in Z and N coordinates, namely, by the transfer of neutrons and protons in the opposite directions. Its time scale 0.5 zs has been extracted from the calculations for the $^{58}$Ni + $^{208}$Pb reaction.

The model~\cite{KarpovSaiko} has been improved by taking into account the equal sharing of the excitation energy and increased fluctuations of the charge asymmetry degree of freedom at the initial stage of a collision. It provides a better agreement of calculated and experimental isotopic distributions of products, especially for the transfer channels of a significant number of protons ($>5$p). It allows us to compare the $^{58}$Ni,$^{64}$Ni + $^{208}$Pb reactions and choose a more appropriate one for production of neutron-rich nuclides. In spite of the increased yields of mass-asymmetric products in the $^{58}$Ni + $^{208}$Pb reaction, the production cross sections of neutron-rich isotopes are larger in the $^{64}$Ni + $^{208}$Pb reaction.

The work has been supported by the RSF Grant No. 19-42-02014.

%

\begin{thebibliography}{}
%
%
\bibitem{Frobrich}
P. Fröbrich, I.I. Gontchar, Phys. Rep. \textbf{292}, 131 (1998)

\bibitem{Karpov}
A.V. Karpov, P.N. Nadtochy, D.V. Vanin, and G.D. Adeev, Phys. Rev. C \textbf{63}, 054610 (2001)

\bibitem{Zagrebaev}
V. Zagrebaev, W. Greiner, J. Phys. G \textbf{31}, 825 (2005)

\bibitem{KarpovSaiko}
A.V. Karpov, V.V. Saiko, Phys. Rev. C \textbf{96}, 024618 (2017)

\bibitem{SaikoKarpov}
V.V. Saiko, A.V. Karpov, Phys. Rev. C \textbf{99}, 014613 (2019)

\bibitem{ZagrebaevKarpov}
V. Zagrebaev, A. Karpov, Y. Aritomo, M. Naumenko, and W. Greiner, Phys. Part. Nucl. \textbf{38}, 469 (2007)

\bibitem{MaruhnGreiner}
J. Maruhn, W. Greiner, Z. Phys. A \textbf{251}, 431 (1972)

\bibitem{WW}
K.T.R. Davies, A.J. Sierk, and J.R. Nix, Phys. Rev. C \textbf{13}, 2385 (1976)

\bibitem{one-body}
A. J. Sierk, J. R. Nix, Phys. Rev. C \textbf{21}, 982 (1980)

\bibitem{stat}
A.V. Karpov, A.S. Denikin, A.P. Alekseev, et al., Phys. At. Nucl. \textbf{79}, 749 (2016); http://nrv.jinr.ru

\bibitem{GEF}
K.-H. Schmidt, B. Jurado, C. Amouroux, and C. Schmitt, Nucl. Data Sheets \textbf{131}, 107 (2016); http://www.khserzhausen.de/GEF.html.

\bibitem{SzilnerCa}
S. Szilner, L. Corradi, G. Pollarolo, et al., Phys. Rev. C \textbf{71}, 044610 (2005)

\bibitem{CorradiNi}
L. Corradi, A.M. Vinodkumar, A.M. Stefanini, et al., Phys. Rev. C \textbf{66}, 024606 (2002)

\bibitem{Awes}
T.C. Awes, R.I.Ferguson, R. Novotny, et al., Phys. Rev. Lett. \textbf{52}, 251 (1984)

\bibitem{Vandenbosch}
R. Vandenbosch, A. Lazzarini, D. Leach, et al., Phys. Rev. Lett. \textbf{52}, 1964 (1984)

\bibitem{Umar}
A.S. Umar, C. Simenel, and W. Ye, Phys. Rev. C \textbf{96}, 024625 (2017)

\bibitem{SapottaNi}
K. Sapotta, R. Bass, V. Hartmann, et al., Phys. Rev. C \textbf{31}, 1297 (1985)

\bibitem{KrolasNi58}
W. Królas, R. Broda, B. Fornal, et al., Nucl. Phys. A \textbf{832}, 170 (2010)

\bibitem{KrolasNi64}
W. Królas, R. Broda, B. Fornal, et al., Nucl. Phys. A \textbf{724}, 289 (2003)

\end{thebibliography}
%
%

\end{document}